# Study of Graph Theory, Distributed Average Consensus Algorithm and Centralized Algorithm

Shen Zheng, Wenzhou Kean University

## 1. Introduction

Graph theory is an important research topic in mathematics, computer science and electrical engineering. There is an extensive amount of research about graph theory about its theoretical framework and its corresponding applications. However, few studies have attempted to relate graph theory with the consensus algorithm, which reveals the graph's underlying structure and numerical stability using the distributed average. The distributed average scheme is vital because we can conclude how the information from one node will evolve in the transmission process. In this paper, we hope to bring closer graph theory and consensus algorithms. Firstly, we give a brief introduction to graph theory by listing a concise definition. Then we analyze and visualize some commonly used graphs. Secondly, we list an overview of the consensus algorithm, including its structure and significance. We then visualize and analyze how the distributed average scheme will reach consensus node values for different probability distributions. Thirdly, we introduce a centralized algorithm about its organizations and functions. Using nodes with different distributions, we analyze how a centralized algorithm will use those nodes to affect the initialize ones. Fourthly, we bring several root-finding algorithms as a supplement to this paper. Finally, we derive conclusions and list important mathematical deviations and algorithms in the Appendix.

The rest of the paper is organized as below. Section 2 introduces graph theory. Section 3 is about consensus algorithm. Section 4 includes some extensions. Section 5 brings the conclusion.

## 2. Graph Theory

### 2.1 Definitions

In mathematics, and especially graph theory, a graph is a data structure for modelling pairwise relationships between objects. A graph consists of vertices and edges. The vertices are mathematical abstractions of objects, whereas the edges are connection lines between different vertices. For example, vertices could represent different tourist attractions and the edges could represent the routes between different tourist attractions.

### 2.2 Directed Graph

Directed Graph is a graph where all the edges are directed from one vertex to another. Following is a simple directed graph where three vertices are connected with an arrow to each other.

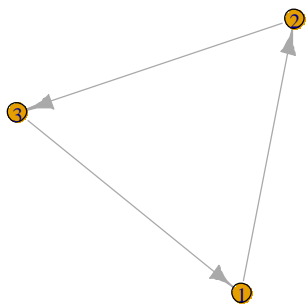

### 2.3 Undirected Graph

Undirected Graph is a graph where all the edges are bidirectional to each other by a line. Following is a simple bidirectional graph where three vertices are connected by lines without arrow.

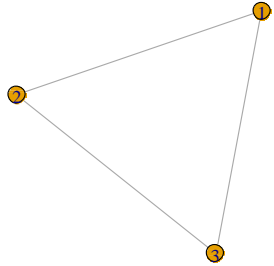

### 2.4 Weighted Graph

A weighted graph is a graph where each branch has a numerical weight. The weight of the edge could represent cost, distance, capacity etc. The first weighted graph comes from a 6*6 adjacency matrix as below.

$$m = \begin{bmatrix} 0 & 0 & 0 & 0 & 0.051197 & 1.34316 \\ 0 & 0 & -0.60881 & 0.40170 & 0 & 0.61322 \\ 0 & -0.60881 & 0 & 0 & -0.63295 & 0 \\ 0 & 0.40170 & 0 & 0 & -0.29831 & 0 \\ 0.051197 & 0 & -0.63295 & -0.29832 & 0 & 0.15625 \\ 1.3432 & 0.61322 & 0 & 0 & 0.15624 & 0 \end{bmatrix}$$

For better visualization, we plot the heatmap of that matrix. The heatmap includes nine nodes from A to F (reordered) with its corresponding values. The darker the colour, the greater the value for the matrix. We could observe from the heatmap plot that A and F has great correlation and, B and D has a significant correlation. This observation is proved by the graph, where A and F, B and D are closely connected.

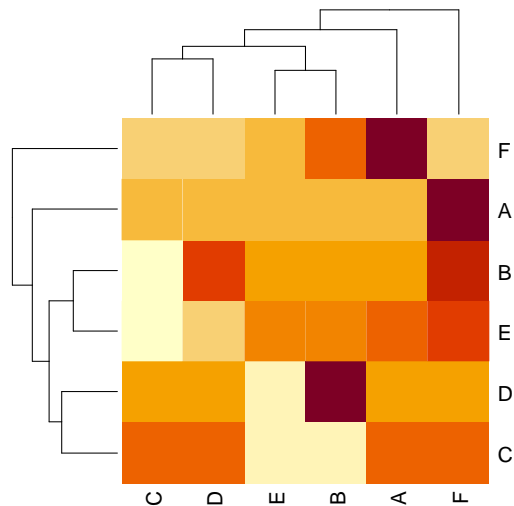

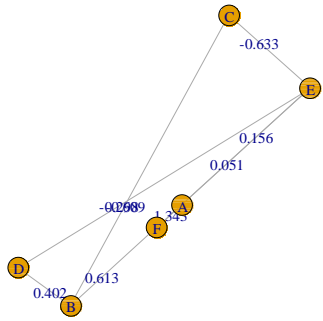

The second weighted graph results from a 5 by 5 adjacency matrix. Each entry in that matrix follow a uniform distribution that has min at -10 and max at 10. Mathematically, we could express a uniform distribution as below, where k is a real number between the min a and the max b.

$$F(k; a, b) = \frac{[k] - a + 1}{b - a + 1}$$

The histogram of that uniform distribution is displayed as below. Note when we move from -10 to 10, the frequency of distribution grows steadily.

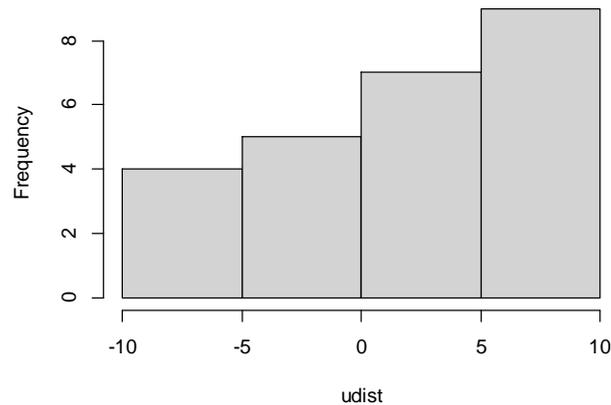

Then we plot the heatmap of that distribution. The heatmap shows that 1,2,3 and 1,4,5 are strongly connected. That observation is demonstrated by the graph where the triangular formed by vertex 1,2,3 and vertex 1,4,5 has significantly bolder edges than others.

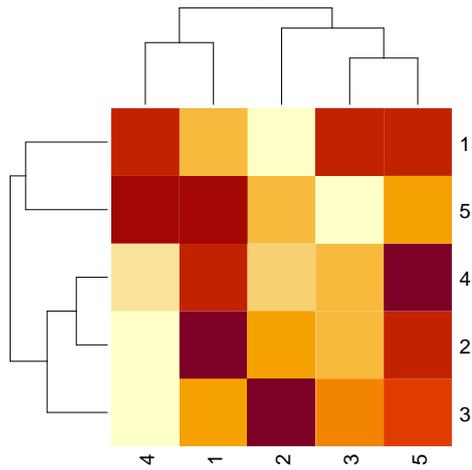

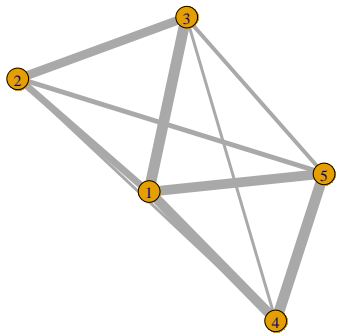

## 2.5 Subgraphs

Subgraph is a graph whose vertex is a subset of another graphs' vertex and whose edges are a subset of another graphs' vertex. In the following example, I first make a ring that has 10 nodes from 1 to 10. Then we plot three subgraphs. The first subgraph includes a chain of nodes from 1 to 5. The second one includes a chain of nodes from 1 to 3. The third one includes a chain of nodes from 1 to 4. Since both vertices and edges of the last three subgraphs are subsets to the first graph, there are all subgraphs of the first one.

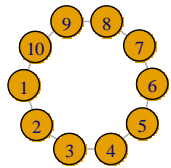
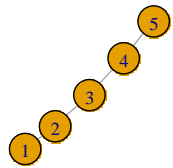

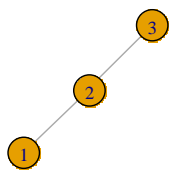
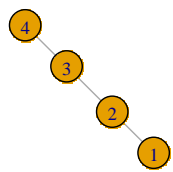

## 2.6 Isomorphism

In graph theory, Isomorphism refers to bijection (one to one correspondence) between vertex sets of one graph A and another graph B such that any two vertices are adjacent in A only if their mappings are adjacent in B.

In the following examples, I show two graphs which has different vertex arrangement but are isomorphic. The left graph has node 5 as the central nodes, which relates to all other nodes, whereas the right graph has node 1 as the central node. However, they are isomorphic graphs.

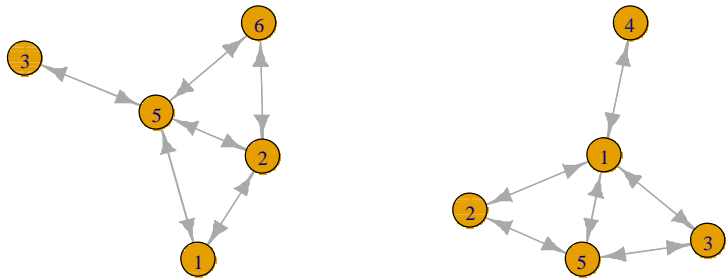

## 2.7 Regular Graph

Regular graph is a graph in which each vertex has the same degree (numbers of neighbors). Note in a multigraph, loops counted twice for degree.

Suppose A is an adjacency matrix of a graph. Then the graph is regular only if an eigenvector of A contains is $j = \begin{bmatrix} 1 \\ ... \\ 1 \end{bmatrix}$. In that case, the eigenvalue will be the degree of the graph and eigenvector from other eigenvalues of that matrix are orthogonal to $j$. According to the Perron-Frobenius theorem, regular graphs of degree k are connected if and only if that eigenvalue k has multiplicity of one.

Following plots shows a group of regular graphs which all have 10 nodes. In the left upper plot, each vertex has a degree of 1. However, the nodes appear in pairs rather than a connected system. In the right upper plot, each vertex has a degree of 2. Besides, all nodes connect to form a ring shape. Each vertex in the left lower plot has a degree of 3, whereas each vertex in the right lower plot has a degree of 4. We can see from these series of regular graphs that when the degree of graphs becomes higher, each node has more connections with each other.

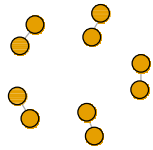
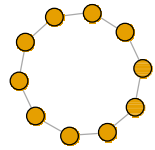
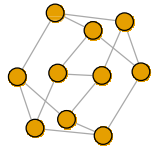
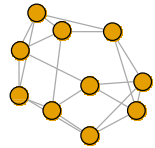

## 2.8 Connected Graph

A connected graph is a graph that has a path from any node to any other nodes. Let's consider the following example, where I show two connected graphs. The left graph has 2 as the centre. If I walk from any points, we will arrive or go through node 2. The right graph has no centre. However, the nodes form a diamond shape where it is possible to walk from any node to others.

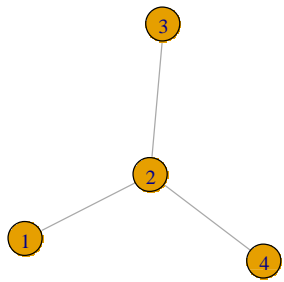
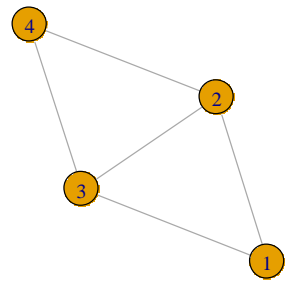

## 2.9 Complete Graph

Complete graph is a graph where each pair of unique nodes connects a unique edge. A complete graph always has $n$ vertices and $\frac{n(n-1)}{2}$ edges. Following are example complete graphs, where there are 3,4,5,6 nodes, respectively. We notice that every node is directly linked to each other. Furthermore, the graph is symmetric.

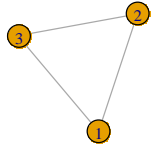 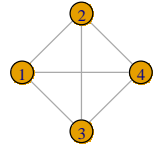

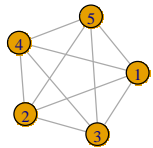 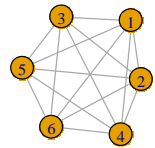

**2.10 Random Graph**

Random Graph is a graph generated by a stochastic process. In a random graph, both the number of graph vertices, edges and their connections are determined by a probability distribution. In the following example, we use the Erdős–Rényi model to generate two different random networks. We choose the $G(n,p)$ model so that we build a graph by connecting nodes randomly. Each edge is included in the graph with a probability independent from every other edge. Each graph has 500 nodes. The left graph has 0.05 the probability for drawing an edge between two arbitrary vertices, whereas the right graph has 0.03 for the same task. We could see from the plot that the left graph has denser connections than the right graph. However, the connections look chaotic (stochastic) instead of organized (definite).

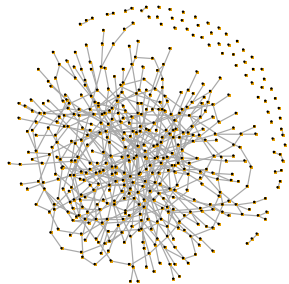 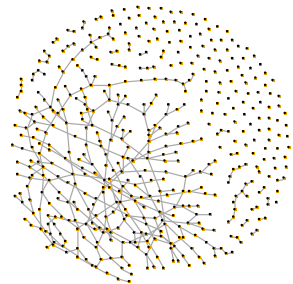

**2.11 Strongly Connected Digraph**

A strongly connected digraph is a directed graph where every vertex is reachable from any other vertex in the same graph. The stronger connected components of a direct graph usually partition into subgraphs that are strongly connected themselves. Following is an example, where we use the Erdős–Rényi model to build a random network with where each vertex has a probability of 0.04 to connect to any other nodes. The separate strongly connected graphs are circled and highlighted in different colours. We could tell from the plot that each partition of a strongly connected digraph is "walkable" from one node to any other.

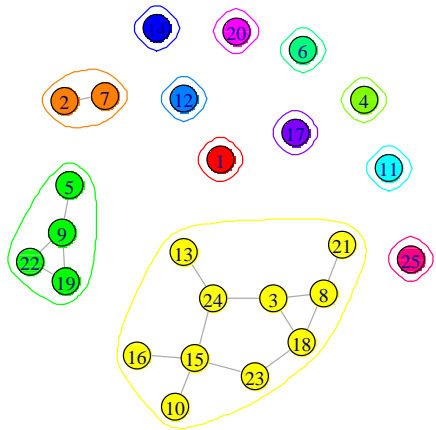

## 2.12 Degree

The degree sum formula, or the handshaking lemma, states that, given a graph $G = (V, E)$, we have the total degree of that graph as:

$$\sum_{v \in V} \deg(v) = 2|E|$$

From the formula, we know that any the number of vertices of an undirected graph with odd degree is even. In the following example, we consider a simple graph with even degree. The graph has six nodes which form a ring. Each node has two neighbours. Therefore, the degree of that graph is 2 for each node.

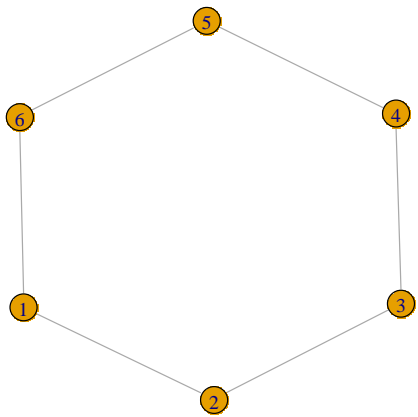

## 2.13 Adjacency Matrix

The adjacency matrix is a matrix used to represent a finite graph. Adjacency matrix must be a square matrix, which means it has the same numbers of rows and columns. The cells of the adjacency matrix show that if pairs of vertices are connected in the graph. If the value of a cell is 1, there is a connection between two specific vertices and otherwise not. Both the row sum and the column sum represent the degree. Besides, an adjacency matrix is always symmetric for an undirected graph. In the following example, we build a directed graph from the adjacency matrix. The matrix is:

$$Adj = \begin{bmatrix} 0 & 1 & 1 & 0 & 0 \\ 0 & 1 & 1 & 1 & 1 \\ 0 & 1 & 1 & 0 & 0 \\ 1 & 0 & 0 & 0 & 1 \\ 0 & 0 & 1 & 0 & 0 \end{bmatrix}$$

The plot from the adjacency matrix is as below. We analyze the plot from the adjacency matrix row by row. In the adjacency matrix's first row, node A connects to node B and node C. Therefore, there is an arrow pointing from A to B and from A to C. In the second row, node B connects to all nodes except node A. Therefore, there is an arrow pointing from node B to all except node A. Note that there is a circle at node B because node B also point to itself. In the third row, node C points to node B and node C itself. Therefore, there is a directed line from B to C. There is also a circle at C. In the fourth row, node D connects to node A and node F with an arrow. And in the fifth row, node F connects to node C with an arrow.

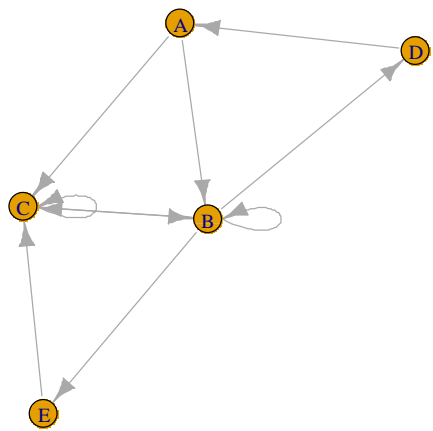

**2.14 Incidence matrix**

Incidence matrix is a matrix that displays the relationship between two classes of objects. The difference between the adjacency matrix and the incident matrix is that incident matrix use rows to represent vertices and columns to represent another entity. In contrast, the adjacency matrix uses both rows and columns to represent the same groups of nodes. Note the incident matrix's cells show that if a vertex connects another vertex in a different entity. If the value of a cell is 1, the vertices connect to the node from other entity and otherwise not. An important aspect of the incidence matrix is that they are not necessarily square matrix. In the following example, we build an incidence matrix. The incidence matrix is:

$$Inc = \begin{bmatrix} 0 & 0 & 0 & 1 & 1 \\ 1 & 0 & 0 & 0 & 1 \\ 1 & 0 & 1 & 1 & 0 \end{bmatrix}$$

The plot from the incidence matrix is as below, where the row shows the first group of the node containing A, B and C, and the column shows the second group of the node containing a, b and c. In the first row, node A connects to node d and node e. Therefore, there is an edge between node A and node d, and node A and node e. In the second row, node B connects to node a and node e. Therefore, there is an edge between node B and node a, and node B and node e. In the third row, node C connects to node a, c and d. Therefore, the id edge between node C and node a, node C and node c, node C and node d. Note that node b is isolated because it is disconnected to any other nodes.

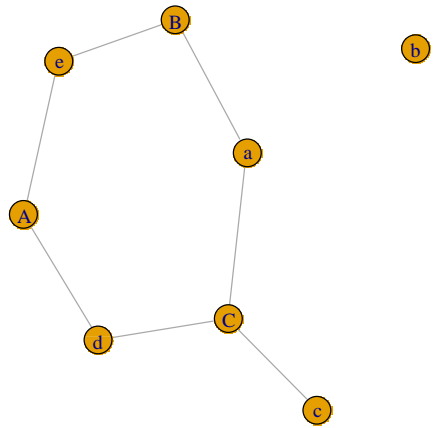

## 2.15 Laplacian Matrix

Laplacian matrix is a special matrix in graph theory. Give a simple graph with n vertices, Laplacian matrix is defined as:

$$L = D - A$$

Where D is the degree matrix and A is the adjacency matrix. Suppose d is the degree. A normalized Laplacian matrix has the elements as:

$$L_{ij}(G) = \begin{cases} 1, & if\ i = j\ and\ d_j \neq 0 \\ -\dfrac{1}{\sqrt{d_i d_j}}, & if\ i\ and\ j\ are\ adjacent \\ 0, & elsewhere \end{cases}$$

In the following example, we use 4 nodes to build a ring and calculate its Laplacian matrix. According to the formula, every cell in the diagonal position is 1. Next, we analyze other cells. In the first row, node 1 connects to all others nodes except node 3. Since the square root of the degree product is 2, If we normalize the output, the cell in the second and the fourth column will have the value -0.5. Others rows have similar behaviours. We note that the Laplacian matrix is symmetric for asymmetric graph and that the values at diagonals and sub diagonals are consistent.

$$L = \begin{bmatrix} 1 & -0.5 & 0 & -0.5 \\ -0.5 & 1 & -0.5 & 0 \\ 0 & -0.5 & 1 & -0.5 \\ -0.5 & 0 & -0.5 & 1 \end{bmatrix}$$

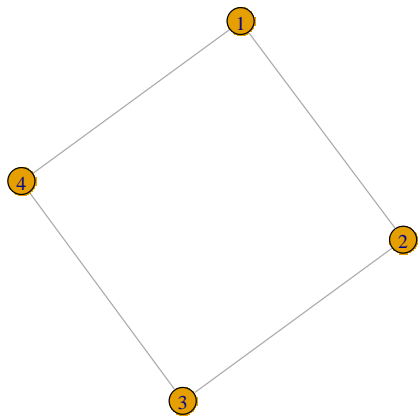

## 2.16 Edge List

Edge list is a data structure in graph theory to represent the graph as a list of its edges. Typically, an edge list will have 2 columns. Each row represents a source value that connects to the target value.

In the following example, we have an edge list with the following source and target columns. The letters represent the node labels.

| Source | Target |
|--------|--------|
| A | B |
| B | B |
| A | C |
| A | D |
| A | F |
| F | A |
| B | E |

We could observe from the matrix that the edges are A—B, B—B, A—C, A—D, A—F, A—F, B—E. Note that node A is both source and target for node F. Therefore, there are 2 edges in the plot between node A and node F. Besides, node B is both source and target for itself. Therefore, there is a circle at node B.

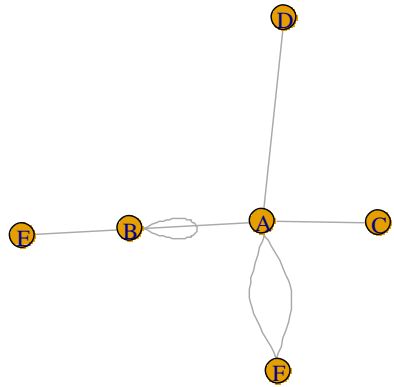

## 3. Consensus Algorithm

In this section, we will give an overview of consensus algorithms and visualize distributed average consensus algorithm on several probability distribution models including uniform distribution, normal distribution, Poisson distribution, binominal distribution, exponential distribution and chi-square distribution.

**3.1 Overview**

A fundamental problem in distributed computing and multi-agent systems is to attain overall system reliability in the presence of several faulty processes. This task often requires coordinating processes to reach consensus, or agree on some data value that is required during computation.

The consensus problem demands agreement among all agents for a single data value. Some of the agents may fail in other ways, so consensus protocols must be fault elastic. The processes must send out their candidate values, communicate with each other and, finally, agree on a unique consensus value.

The consensus problem is fundamental in control of multi-agent systems. One approach to producing consensus is for all agents to agree on a majority value. In this context, a majority requires more than half of valid votes. However, one or more outliers may skew the resultant outcome such that the consensus algorithm fails to converge.

Protocols that solve consensus problems are designed to deal with limited numbers of faulty processes. According to Coulouris [1], A consensus protocol tolerating halting failures must satisfy the following properties.

**Termination**

Eventually, every correct process decides some value.

**Integrity**

If all the correct processes proposed the same value, then any correct process must decide that value.

**Agreement**

Every correct process must agree on the same value.

In the following experiments, we will calculate and visualize the distributed average with Uniform distribution that has 20 nodes according to the following procedures.

(1) We construct an adjacency matrix such that ant two connected nodes in that matrix has the value of 1 and otherwise 0.
(2) We construct and visualize the graph from that adjacency matrix.
(3) We find the number of neighbours of each node using the L1 sum of the adjacency matrix.
(4) We build an averaging process matrix which has the value as one divided by the neighbours of that node if two nodes are connected and else 0.
(5) We establish an iterative scheme such that the consensus node value could be visualized concerning time.
(6) We visualize the distributed averaging result, the histogram of node degree and the cumulative frequency scatter plot.
(7) We interpret the results from different probability distributions.

**3.2 Distributed Average with Uniform Distribution**

In the following example, we build a uniform distribution graph with -1 as the min and 1 as the max. According to the result from uniform distribution's distributed averaging, the consensus algorithm starts at node value between 0 and 1 and converge to the 0.4 within 5 iterations. After that, the consensus value remains stable. The histogram of node degree (numbers of node neighbors) is approximately symmetric with no significant outliers. The cumulative frequency plot demonstrates most node degree centers around 8 to 12. The result show use that most nodes has 8 to 12 neighbors.

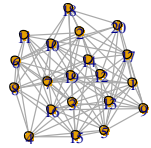
Uniform Distribution

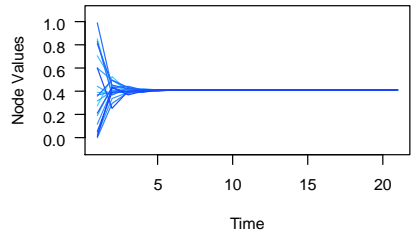
Distributed Averaging

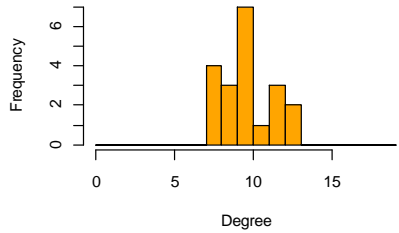
Histogram of node degree

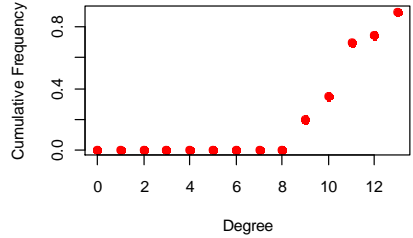
Cumulative Frequency Scatter Plot

### 3.3 Distributed Average with Normal Distribution

In the following example, we build a standard normal distribution model such that the graph has mean of 0 and standard deviation of 1. According to the result from the distributed averaging, the consensus algorithm starts at node value between -3 and 2 and converge to the 0 within 5 iterations. From the $5^{th}$ to the $10^{th}$ iteration, the consensus value has a slight fluctuation. After the $10^{th}$ iteration, the value remains stable at 0. The histogram of node degree is slightly left skewed with no significant outliers. The cumulative frequency plot demonstrates most node degree centers around 8 to 14. The result show use that most nodes has 8 to 14 neighhours.

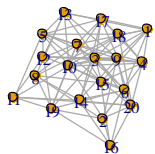
Normal Distribution

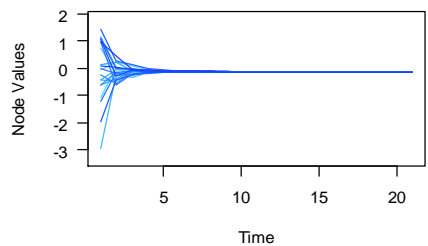
Distributed Averaging

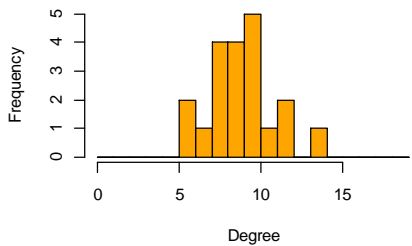
Histogram of node degree

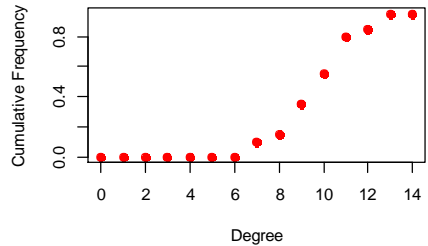
Cumulative Frequency Scatter Plot

### 3.4 Distributed Average with Poisson Distribution

In the following example, we build a default Poisson distribution model. According to the result from the distributed averaging, the consensus algorithm starts at node value between -3 and 2 and converge to the 0 within 5 iterations. From the 5th to the 8th iteration, the consensus value has a slight fluctuation. After the 10th iteration, the value remains stable at 0. The histogram of node degree demonstrates most node degree follows a Poisson distribution. The cumulative frequency scatter plot show use that most nodes has 7 to 15 neighbors.

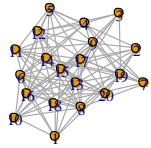

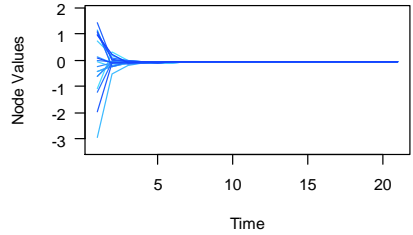

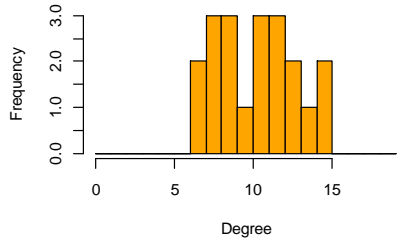

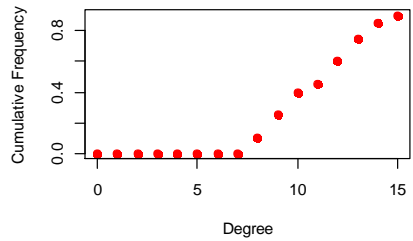

### 3.5 Distributed Average with Binominal Distribution

In the following example, we build a binominal distribution model. According to the result from the distributed averaging, the consensus algorithm starts at node value between -3 and 2 and converge to the 0 within 5 iterations. The histogram of node degree is approximately symmetric. The cumulative frequency plot scatter plot show that most nodes has 6 to 14 neighbors.

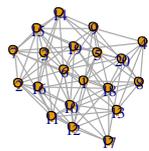

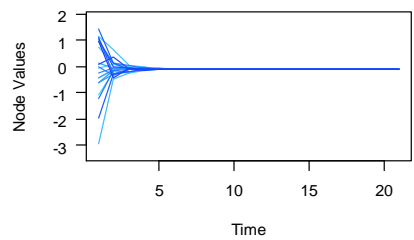

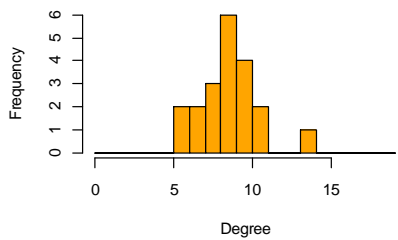

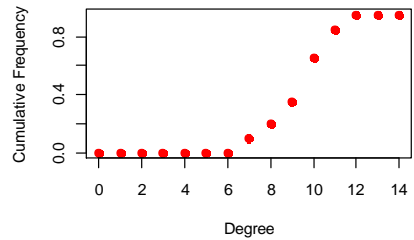

### 3.6 Distributed Average with Exponential Distribution

In the following example, we build an exponential distribution model that has rate of 1.

According to the result from the distributed averaging, the consensus algorithm starts at node value between 0 to 5 and converge to the approximately 1.3 within 5 iterations. It is not certain from the histogram of node degree how the node connects with each other. However, the cumulative frequency plot scatter plot show that most nodes has 6 to 14 neighbors.

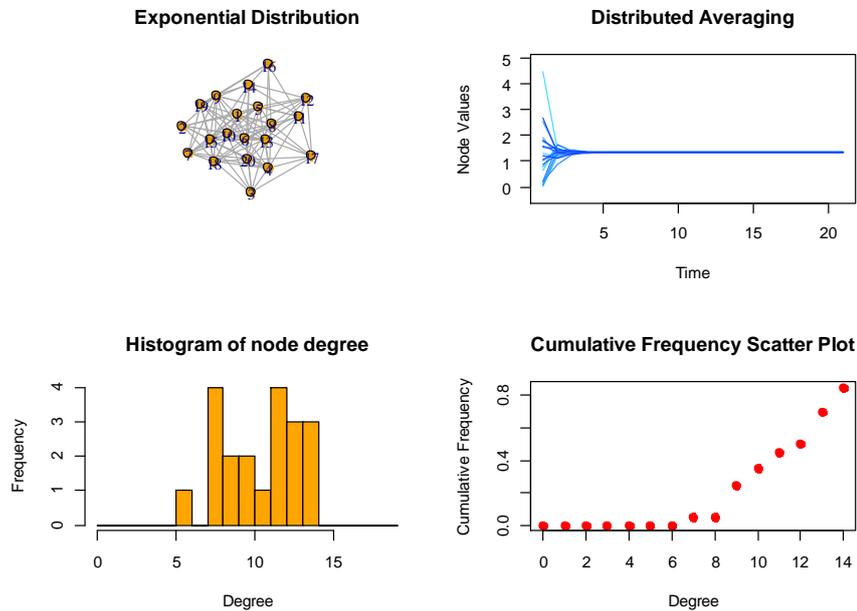

### 3.7 Distributed Average with Chi Square Distribution

In the following example, we build a Chi Square distribution model that has rate of 1. According to the result from the distributed averaging, the consensus algorithm starts at node value between 0 to 5 and converge to the approximately 1.2 within 5 iterations. The histogram of the node degree is approximately symmetric. Besides, the cumulative frequency plot scatter plot show that most nodes has 7 to 16 neighbors.

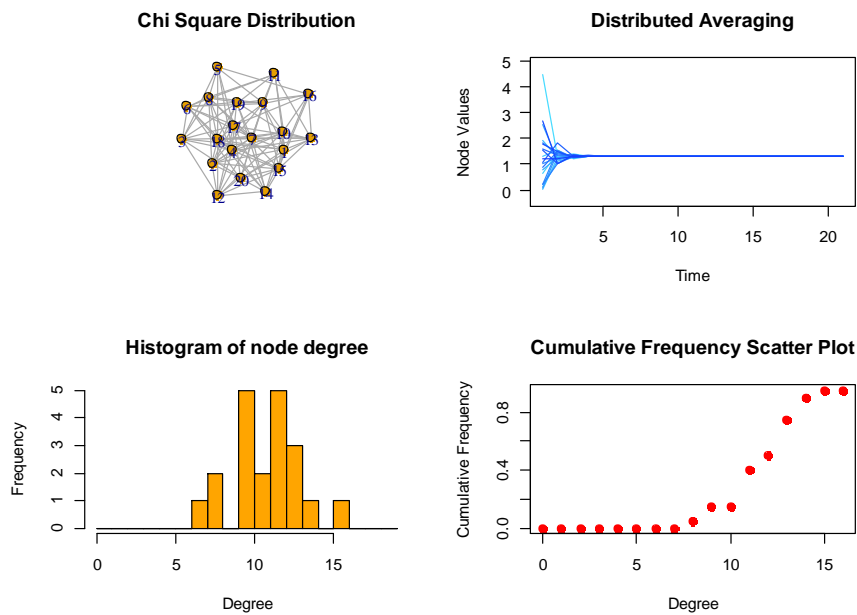

## 4. Centralized Algorithm

In this section, we will study centralized algorithm. We will first give an overview of centralized algorithms. We will then implement and visualize them on several probability distribution models including uniform distribution, normal distribution, Poisson distribution, exponential distribution and chi-square distribution.

**4.1 Overview**

In graph theory, central server is the most essential vertices within a graph. Centralized systems are systems that use server architecture where one or more client nodes are directly connected to a central server. Suppose we enter the search term "US presidential Election" on Google, this search term is sent as a request to the Google serves which then return the pages links based on its recommendation system. In that example, we are the client node, whereas Google is the central server. Note a centralized system must have a central node and a communication link. The problem in real application, especially in numerical computation is that the centra nodes are difficult to acquire. Since centralized system relies on direct transmissions, it is possible that the simulated central points fail to approximate the real center position. In this case, there will be massive accuracy loss from the transmission process. There is a strong need to study the centralized process to reduce the potential accuracy loss.

In the following examples, we discuss how efficient and effective do centralized algorithm converge with different probability distributions. First, we randomize 100 nodes as the starting one. Then we iteratively add 1000 nodes each round. After that, we plot the 2d density plot to show the distribution density of the nodes. The upper left, upper right, lower left, lower right plot shows the node distribution density after 1 iteration, 5 iterations, 10 iterations and 20 iterations, respectively.

**4.2 Centralized Algorithm with Uniform Distribution**

In the following example, I define a uniform distribution that has min of 2 and max of 25. Then I visualize how centralized algorithm with uniform distributed nodes will nodes distributions. According to the plot, after 20 iterations, the nodes become uniformly distributed. There are some disruption points. And there are no distinct central points.

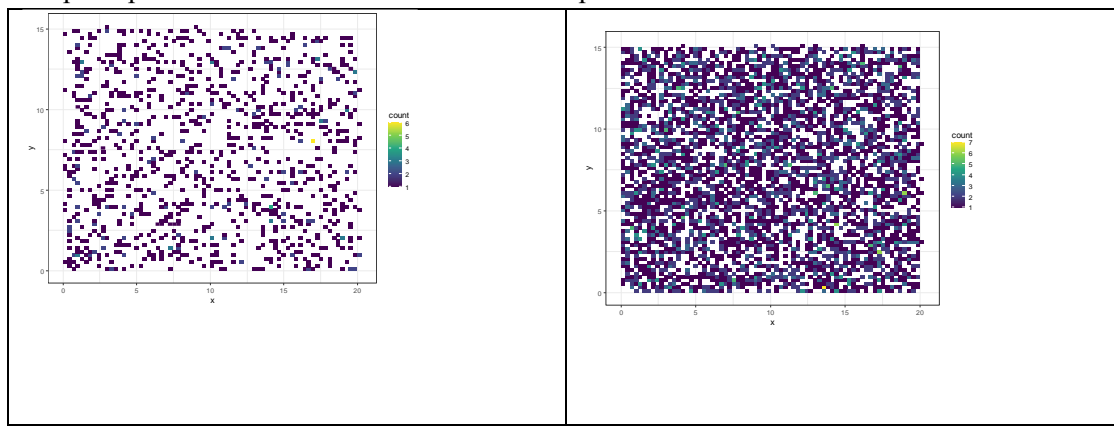

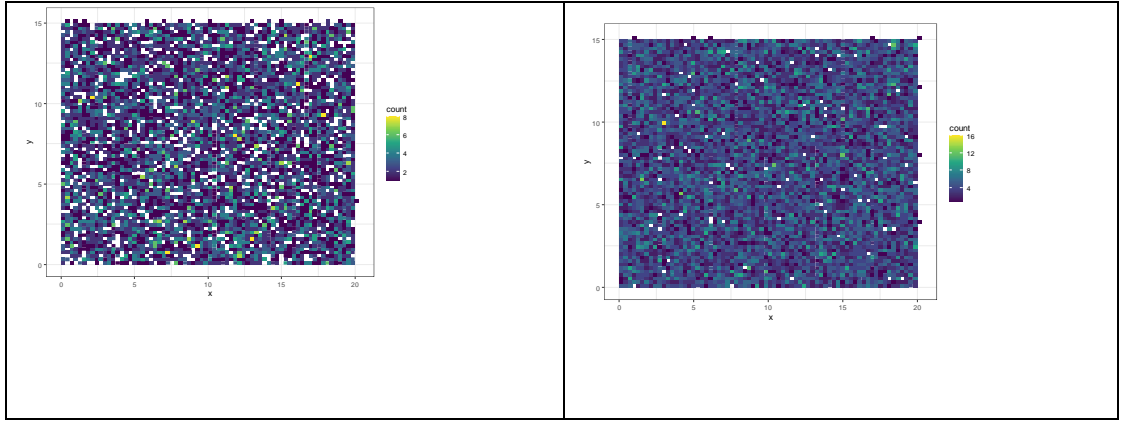

### 4.3 Centralized Algorithm with Normal Distribution

In the following example, I define a normal distribution that has standard deviation of 2, mean of 10 horizontally and of 7.5 vertically so that disruptions in visualization will be clearer. Then I visualize how centralized algorithm with normally distributed nodes will affect nodes distributions. According to the plot, after 20 iterations, the nodes become normally distributed. Although there are significant amounts of outliers outside the central region, the central node is distinct and symmetric.

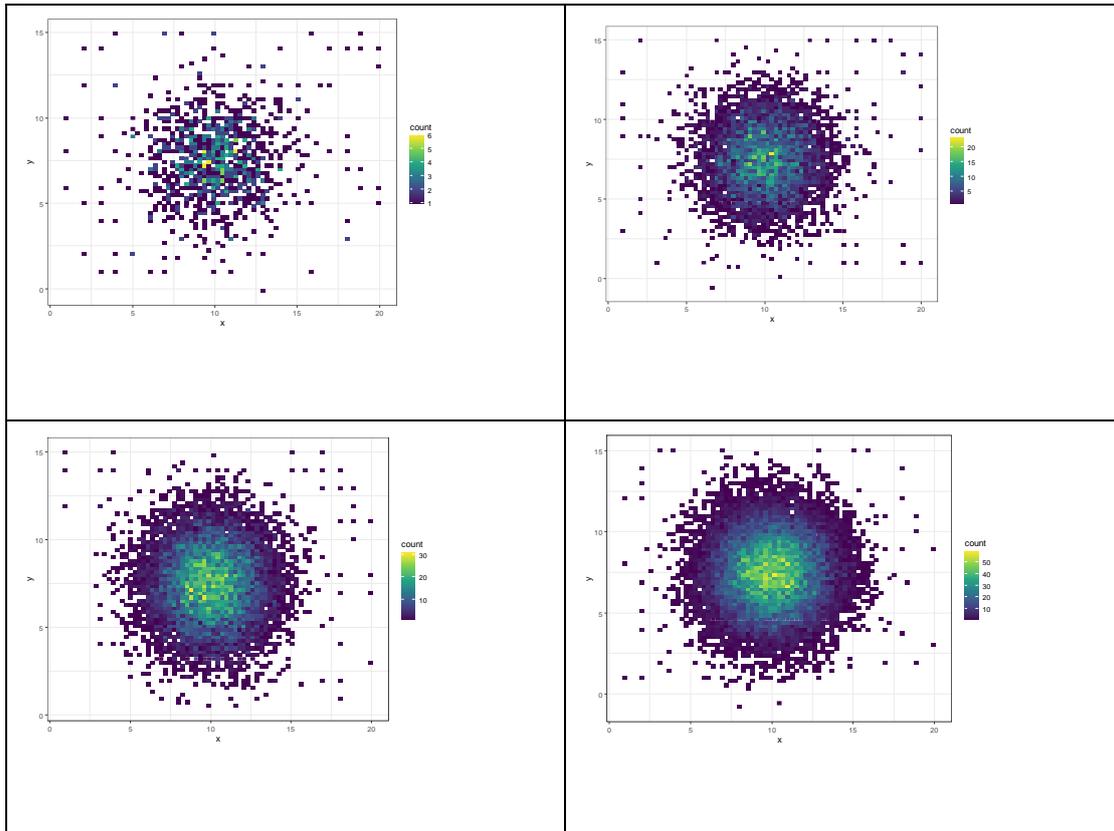

### 4.4 Centralized Algorithm with Poisson Distribution

In the following example, I define a Poisson distribution that has mean of 0 and lambda of 2. Then I visualize how centralized algorithm with Poisson distribution nodes will affect nodes distributions. According to the plot, after 20 iterations, the nodes is near to Poisson distribution. While more nodes close the axes as the iterations goes on, there are few nodes grow far away from the axis. The central region could be seen near the origin. However, the central node is not clear from the plot.

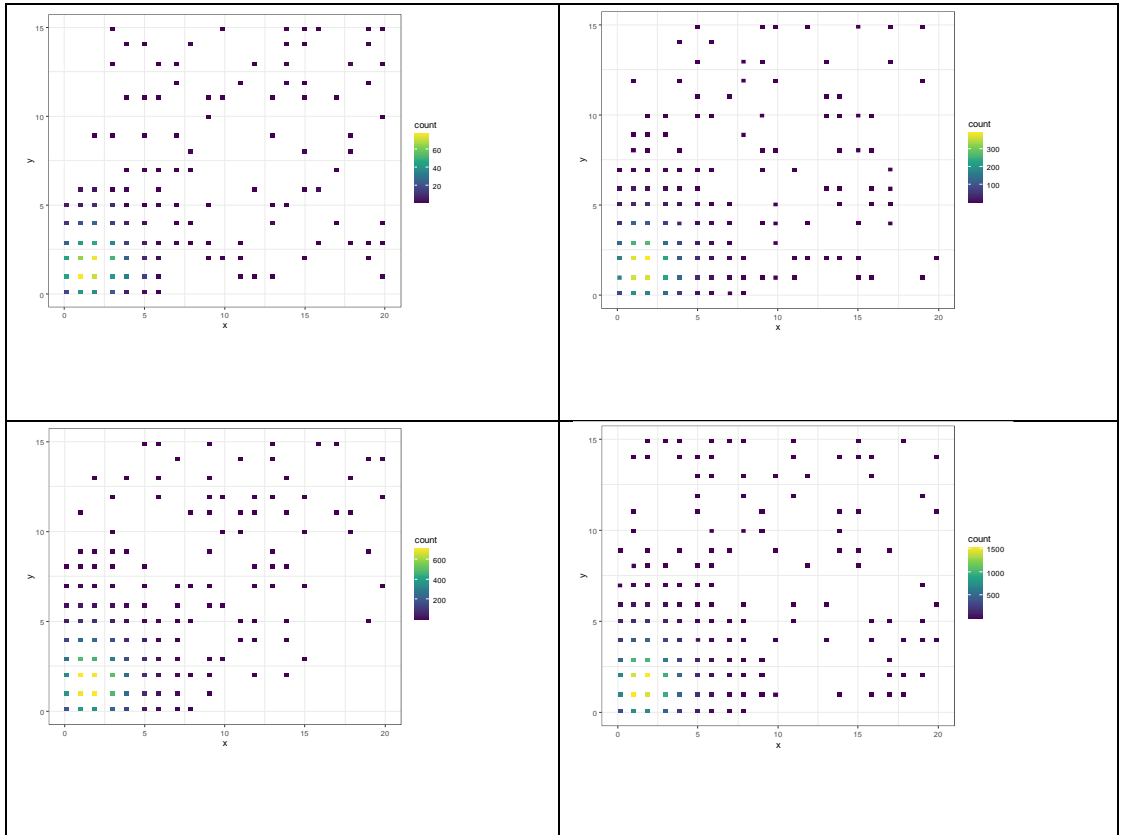

### 4.5 Centralized Algorithm with Exponential Distribution

In the following example, I define an exponential distribution. Then I visualize how centralized algorithm with exponentially distributed nodes will affect nodes distributions. According to the plot, after 20 iterations, the nodes approximately follows exponential distribution. Indeed, there is no distinct central regions because most adding nodes are grouped and overlap with each other. However, the central node, which is the origin point, is distinct.

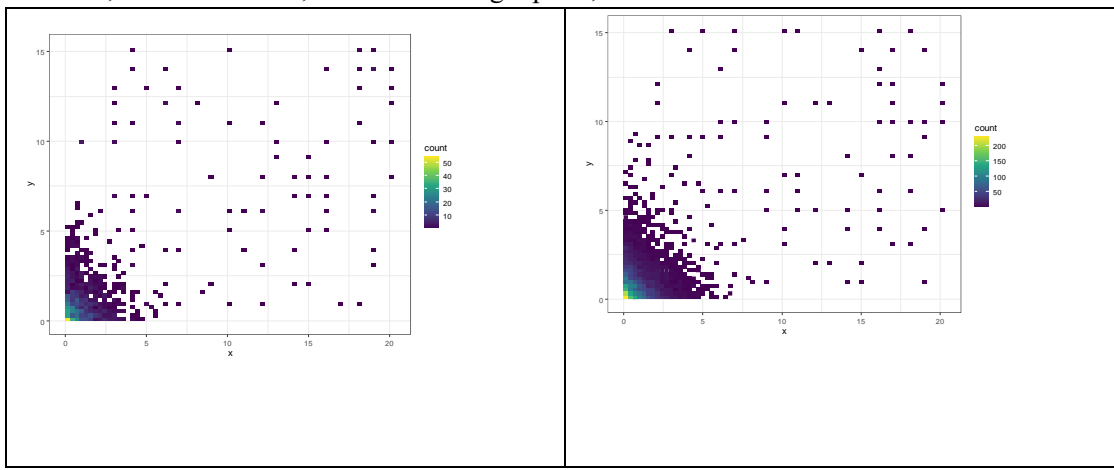

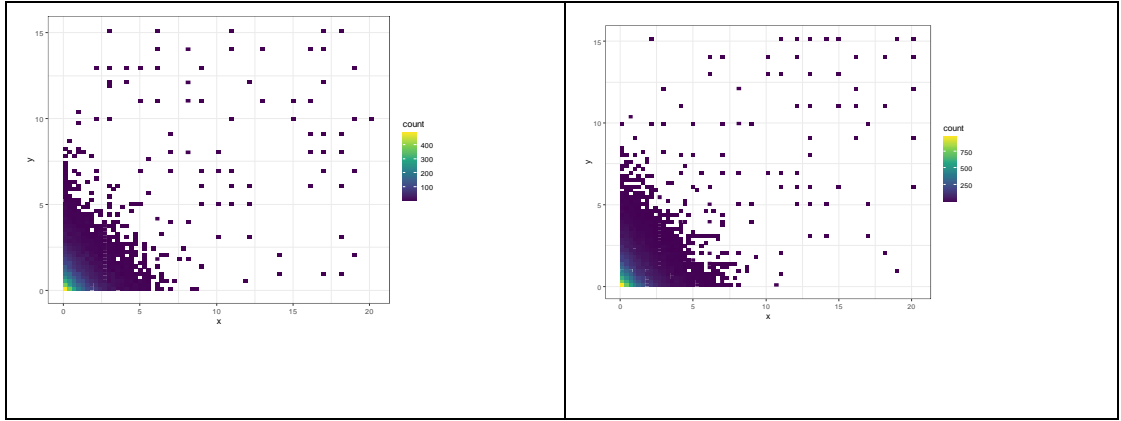

### 4.6 Centralized Algorithm with Chi Square Distribution

In the following example, I define a Chi Square distribution with degree of freedom of 3. Then I visualize how centralized algorithm with Chi Square distribution nodes will affect nodes distributions. According to the plot, after 20 iterations, the nodes approximately follows Chi Square distribution. Both the central region and the central node is clear from the plot.

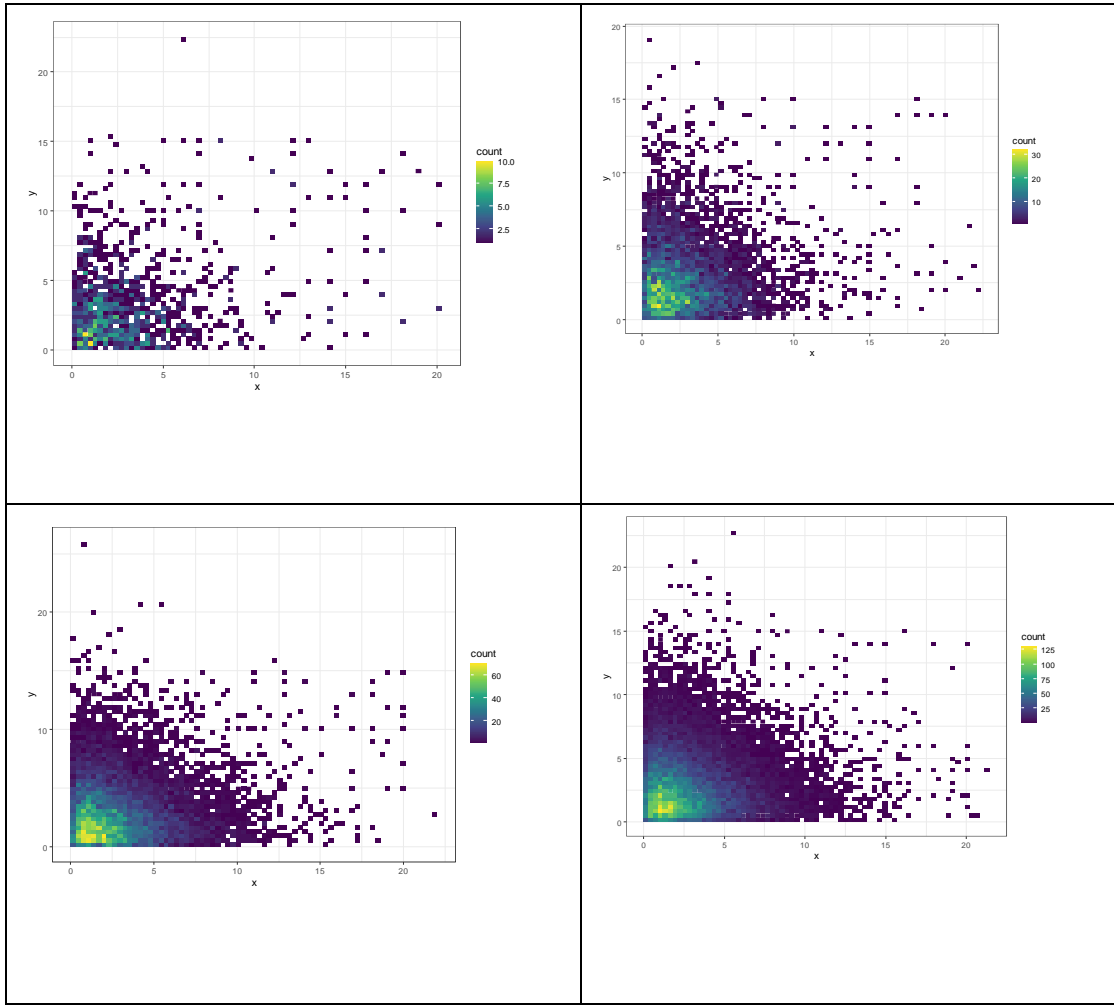

### 5. Root Finding Algorithms

Graphs in social network analysis often come down to finding roots for an optimization result. In this section, we consider a simple example with the following function.

$$f(x) = x * \cos(x) - \sin(x), 0 \leq x \leq 5$$

The plot of the function in the domain is as below. We could observe from the plot that one root is at 0, while another root is between 4 and 5. To find that second root, we utilize four different root finding methods including fixed point method, bisection method, secant method and newton method. The detailed root finding algorithm for those methods are included in the Appendices.

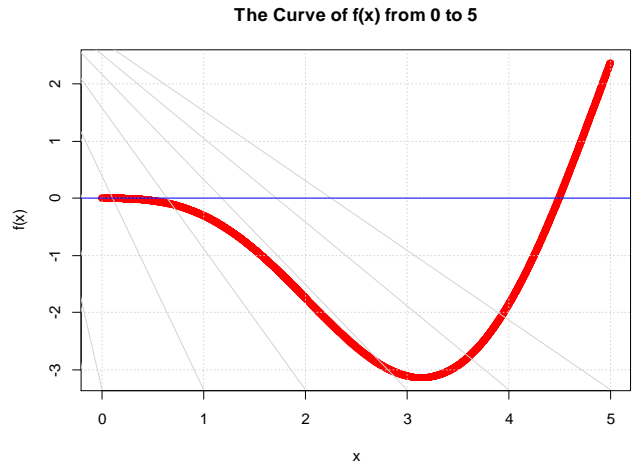

If we implement all these methods in the domain between x = 4 and x = 5, the results of various algorithms are shown as below:

| Root Finding Algorithm Results | | |
|---|---|---|
| **Method** | **Iterations** | **Result** |
| Fixed Point | 3 | 0 |
| Bisection | 4 | 4.493409 |
| Secant | 4 | 4.493409 |
| Newton Raphson | 3 | 4.493409 |

As we could observe from the plot, Newton Raphson method is the most efficient method Newton Raphson finds the accurate root between 4 and 5 within 3 iterations, whereas bisection and secant method has to use 4 iterations for the same result. Although fixed point method also completes the task in 3 iterations, it gets the wrong answer.

## 6. Conclusion

This paper has shown that it is possible to connect graph theory with distributed average consensus algorithms, centralized algorithms, and root finding algorithms. Firstly, we study some fundamental aspects of graph theory. We introduce some basic graph structures such as a directed graph, random graph, and weighted graphs. We also use the adjacency matrix, incidence matrix, and Laplacian matrices to relate matrices with graphs. Secondly, we investigate distributed average consensus algorithm on various probability distributions such as normal distribution and Poisson distribution. Our results show that exponentially distributed nodes have the fastest convergence on consensus node value, where normally distributed nodes have the most stable consensus node value. Thirdly, we examine the centralized algorithm on the same groups of probability distributions. Our findings show that the centralized algorithm is the most robust and efficient when the additive nodes follow normal or Chi-Square distribution. Finally, we discuss root-finding algorithms. Among the four chosen algorithms, Newton-Raphson methods outperform others in optimization efficiency. Future works are encouraged to relate graph theory with the decentralized algorithm, which is more stable and efficient for big data.

# Appendices
## 1. Matrices and Graphs
### 1.1 Adjacency Matrices

The adjacency matrix of a graph has entries determined by the following equation

$$A_{i,j} = \begin{cases} 1, & \text{if vertex } v_i \text{ is adjacent to vertex } v_j \\ 0, & \text{otherwise} \end{cases}$$

**Corollary 1:**

*Let G be a graph with adjacency matrix A and k be a positive integer. The sum $S_n$ is defined by:*

$$S_n = \sum_{i=1}^{k} A^i$$

$S_n$ is a n * n matrix whose entries counts the number of paths that has length less than or equal to k between vertices $v_i$ and $v_j$

### 1.2 Incidence Matrices

The incident matrix of a graph has entries determined by the following equation

$$Q_{i,j} = \begin{cases} 1, & \text{if vertex } v_i \text{ is incident to edge } e_j \\ 0, & \text{otherwise} \end{cases}$$

### 1.3 Laplacian Matrices

The Laplacian matrix of a graph has entries determined by the following equation

$$L_{i,j} = \begin{cases} -1, & \text{if vertex } v_i \text{ is adjacent to vertex } v_j \\ 0, & \text{if vertex } v_i \text{ is not adjacent to vertex } v_j \\ d_i, & \text{if } i = j \end{cases}$$

If we define a D(G) to be a diagonal matrix whose entries are degrees of each vertex. And we define A(g)

as the adjacency matrix. The Laplacian matrix of a graph G could be defined as:
$$L(G) = D(G) - A(G)$$
Laplacian matrix can also be written in terms of incidence matrix Q(G) as below:
$$L(G) = Q(G)Q^T(G)$$

## 2. Mathematical Formulation for Probability Distributions
### 2.1 Uniform Distribution
(1) Density Function

$$f(x) = \begin{cases} \dfrac{1}{b-a}, a \leq x \leq b \\ 0, elsewhere \end{cases}$$

(2) Mean

$$\mu(x) = \frac{a+b}{2}$$

(3) Variance

$$\sigma^2(x) = \frac{(b-a)^2}{12}$$

### 2.2 Normal Distribution
(1) Density Function

$$n(x;\mu,\sigma) = \frac{1}{\sqrt{2\pi}\sigma} e^{-\frac{1}{2\sigma^2}(x-\mu)^2}, -\infty < x < \infty$$

(2) Mean

To obtain the mean, we first calculate

$$E[x-\mu] = \int_{-\infty}^{\infty} \frac{x-\mu}{\sqrt{2\pi}\sigma} e^{-\frac{1}{2\sigma^2}(x-\mu)^2} dx$$

Setting $z = \frac{x-\mu}{\sigma}$ and $dx = \sigma dz$, we have

$$E[x-\mu] = \frac{1}{\sqrt{2\pi}} \int_{-\infty}^{\infty} z\, e^{-\frac{1}{2}z^2} dz = 0$$

Therefore

$$E[x] = E[\mu] = \mu$$

Therefore

$$\mu(x) = \mu$$

(3) Variance

$$E[(x-\mu)^2] = \frac{1}{\sqrt{2\pi}\sigma} \int_{-\infty}^{\infty} (x-\mu)^2 e^{-\frac{1}{2}(\frac{x-\mu}{\sigma})^2} dx$$

Setting $z = \frac{x-\mu}{\sigma}$ and $dx = \sigma dz$, we have

$$E[(x-\mu)^2] = \frac{\sigma^2}{\sqrt{2\pi}} \int_{-\infty}^{\infty} z^2 e^{-\frac{1}{2}z^2} dz$$

Integration by parts with $u = z$ and $dv = ze^{-\frac{z^2}{2}}$ so that $du = dz$ and $v = -e^{-\frac{z^2}{2}}$, we get

$$E[(x-\mu)^2] = \sigma^2(0+1) = \sigma^2$$

Therefore

$$\sigma^2(x) = \sigma^2$$

**2.3 Poisson Distribution**

(1) Density Function

$$p(x; \lambda t) = \frac{e^{-\lambda t}(\lambda t)^x}{x!}, x = 0,1,2,3,\ldots$$

(2) Mean

$$\mu(x; \lambda t) = \lambda t$$

(3) Variance

$$\sigma^2(x; \lambda t) = \lambda t$$

**2.4 Binominal Distribution**

(1) Density Function

$$b(x; n, p) = \binom{n}{x} p^x (1-p)^{n-x}, x = 0,1,2,3,\ldots,n$$

(2) Mean

Let the outcome on the $j^{th}$ trial be represented by a Bernoulli random variable $I_j$ which assumes the values 0 and 1 with probabilities q and p, respectively. Therefore, in a binomial experiment the number of successes can be written as the sum of the n independent indicator variables. Hence,

$$X = I_1 + I_2 + \cdots + I_n$$

We note that the mean of any $I_j$ is $E[I_j] = 0 * (1-p) + 1 * p = p$. Therefore, we have the mean of the binomial distribution as

$$\mu(x; n, p) = E[x] = \sum_{i=1}^{n} E(I_i) = np$$

(3) Variance

The variance of any $I_j$ is $\sigma_{I_j}^2 = E(I_j) - p^2 = 0^2 * (1-p) + 1^2 * p - p^2 = p(1-p)$

We state the following corollary.

**Corollary 2:**

If $X_1, X_2, \ldots X_n$ are independent random variables, then

$$\sigma_{\sum_{i=1}^{n} a_i X_i}^2 = \sum_{i=1}^{n} a_i^2 \sigma_{X_i}^2$$

According to Corollary 2, for the case of n independent Bernoulli variables. We have the variance of the binomial distribution as

$$\sigma^2(x; n, p) = \sum_{i=1}^{n} \sigma_{I_i}^2 = \sum_{i=1}^{n} p(1-p) = np(1-p)$$

**2.5 Exponential Distribution**

(1) Density Function

$$f(x; \beta) = \begin{cases} \frac{1}{\beta} e^{-\frac{x}{\beta}}, x > 0 \\ 0, elsewhere \end{cases}, where\ \beta > 0$$

(2) Mean

$$\mu(x; \beta) = \beta$$

(3) Variance

$$\sigma^2(x;\beta) = \beta^2$$

**2.6 Chi-Square Distribution**

(1) Density Function

$$f(x;v) = \begin{cases} \dfrac{1}{2^{\frac{v}{2}}\tau(\frac{v}{2})} x^{\frac{v}{2}-1} e^{-\frac{x}{2}}, x > 0 \\ 0, elsewhere \end{cases}, where\ v\ is\ a\ postive\ integer$$

(2) Mean

$$\mu(x;v) = v$$

(3) Variance

$$\sigma^2(x;v) = 2v$$

# 3. Pseudocode for Root Finding Algorithms
## 3.1 Fixed point method

**Algorithm 1: Fixed Point Iteration**

**Require**: initial approximation $p_0$; tolerance TOL; maximum number of iterations $N_0$
Set $i = 1$
While $(i \leq N_0)$:
 $p = g(p_0)$
 If $|p - p_0| < TOL$:
  Return $p$
  break
 $i += 1$
 $p_0 = p$
**Return** "fail to converge"

## 3.2 Bisection method

**Algorithm 2: Bisection Iteration**

**Require**: end points $a, b$; tolerance TOL; maximum number of iterations $N_0$
Set $i = 1; FA = f(a)$
While $(i \leq N_0)$:

 $p = a + \dfrac{b-a}{2}$

 $FP = f(p)$

 If $FP = 0$ or $\dfrac{b-a}{2} < TOL$:

  Return $p$
  break
 $i += 1$
 If $FA * FP > 0$:
  $a = p$
 Else:
  $b = p$
**Return** "fail to converge"

## 3.3 Secant method

**Algorithm 3: Secant Iteration**

**Require**: Initial approximations $p_0, p_1$; tolerance TOL; maximum number of iterations $N_0$

Set $i = 2; q_0 = f(p_0); q_1 = f(p_1)$

While $(i \leq N_0)$:

$p = p_1 - q_1 * \frac{p_1 - p_0}{q_1 - q_0}$

If $|p - p_1| < TOL$:

**Return** $p$

break

$i \mathrel{+}= 1$

$p_0 = p_1$

$q_0 = q_1$

$p_1 = p$

$q_1 = f(p)$

**Return** "fail to converge"

## 4. R code for implementation